\documentclass{aastex62}
\graphicspath{{./}{figures/}}

\shorttitle{Sample article}
\shortauthors{Estrela et al.}

\begin{document}

\title{Surface and oceanic habitability of Trappist-1 planets under the impact of flares}

\correspondingauthor{Raissa Estrela}
\email{restrela@jpl.nasa.gov}

\author{Raissa Estrela}
\affil{Jet Propulsion Laboratory, California Institute of Technology \\
4800 Oak Grove Dr, Pasadena, CA 91109}
\affiliation{Center for Radioastronomy and Astrophysics Mackenzie, Rua da Consolacao, Sao Paulo, Brazil}

\author{Sourav Palit}
\affiliation{Department of Physics, Indian Institute of Technology Bombay (IITB), Mumbai, India 400076}

\author{Adriana Valio}
\affiliation{Center for Radioastronomy and Astrophysics Mackenzie, Rua da Consolacao, Sao Paulo, Brazil}



\begin{abstract}
 The discovery of potentially habitable planets around the ultracool dwarf star Trappist-1 naturally poses the question: could Trappist-1 planets be home to life? These planets orbit very close to the host star and are most susceptible to the UV radiation emitted by the intense and frequent flares of Trappist-1. Here we calculate the UV spectra (100 - 450 nm) of a superflare observed on Trappist-1 with the K2 mission. We couple radiative transfer models to this spectra to estimate the UV surface flux on planets in the habitable zone of Trappist-1 (planets $e$, $f$, and $g$), assuming atmospheric scenarios based on a pre-biotic and an oxygenic atmosphere. We quantify the impact of the UV radiation on living organisms on the surface and on a hypothetical planet ocean. Finally, we find that for non-oxygenic planets, UV resistant lifeforms would survive on the surface of planets f and g. Nevertheless, more fragile organisms (i.e. \textit{E. coli}) could be protected from the hazardous UV effects at ocean depths greater than 8m. If the planets have an ozone layer, any lifeforms studied here would survive in the HZ planets.

\end{abstract}




\section{Introduction}

The Trappist-1 system is a key future target in the search for life. This system contains seven Earth-sized planets orbiting around a M-dwarf, with three of them (Trappist-1 e, f and g) located in the Habitable Zone (HZ) of the host star, a region with the right conditions to harbour life \citep{Gillon2017, Wolf17, Grimm2018, Papa18}. In particular, Trappist-1e seems to be the most favourable to have a habitable temperature of 279k under a clear atmosphere \citep{Linco18}. Since M dwarf stars spend $\sim$ 10$^{10}$ years in the main sequence there is enough time for complex forms of life to develop and evolve \citep{Line04, Scalo07}. However, a recent study from \cite{Vida2017} using Kepler/K2 data has found frequent flaring of the Trappist-1 star, bringing the habitability of these worlds into question. A total of 42 flares with energy of 10$^{30}$-10$^{33}$ erg was detected, with an average time interval of 28 hours between consecutive flares. The strongest flare found, released an energy of 1.24 $\times$ 10$^{33}$ erg,  similar energy range to the largest solar flare in history, namely the ‘Carrington event’ in 1859. These energetic flares are also called superflares, and have been found in other M dwarf stars like DG CVn \citep{Drake14}, AD Leo \citep{Hawley1991}, and very recently in Proxima Centuri \citep{Howard18, MacG18, Loyd18}, among others.

Superflares can release significant amount of ultraviolet radiation in XUV, EUV and FUV, which affect the planetary atmosphere in many different ways, such as atmospheric mass loss, modification of chemical composition over a wide range of altitude, and instability of various atmospheric layers. Moreover accelerated energetic protons associated with such strong flares may produce copious amount of odd nitrogen and odd hydrogen in the upper stratosphere and mesosphere, that can potentially destroy the ozone layer (\cite{Segura2010}). On Earth, the ozone layer is responsible for absorbing most of the solar ultraviolet radiation arriving at our planet. In particular, the radiation that is the most threatening for life, like UVC (100-280 nm), is completely absorbed by the ozone layer, while UVB (280-315 nm) has an absorption of 95\% and UVA (315-400nm) can reach the Earth`s surface. Ozone can also absorb the stellar XUV emission, which is particularly important for the Trappist-1 system as the XUV doses received at the top of the atmosphere (TOA) of these planets are higher ($\sim$ 65 times for planet-e) than those at Earth \citep{Pea19,Yam19}. Therefore, the ozone layer acts as a shield that protects complex lifeforms living on the surface of our planet from the harmful ultraviolet radiation. Other molecules, such as H$_{2}$O and CO$_{2}$, also absorb wavelengths shorter than 200nm even in anoxic atmospheres.

Particularly for the three HZ planets of Trappist-1, the effects of strong flares/superflares and associated energetic particle events could be even worst, as they orbit very close to the host star (0.029, 0.037 and 0.0451 AU, respectively). This would probably put the habitability and atmospheric stability of these planets at great risk. On the other hand, if these planets still host an atmosphere, depending on its composition, this atmospheric layer could probably attenuate the UV radiation from the host star and allow surface lifeforms to develop.

However, the composition of the atmosphere of Trappist-1's planets is still not known. An attempt to investigate the atmosphere of these planets was recently performed by \cite{DeWit2018}. They claim that Trappist-1 planets have an atmosphere with a lack of hydrogen, which supports that these planets are more terrestrial like, with an exception for Trappist-1 g, for which a hydrogen atmosphere can not be excluded. Therefore, different atmospheric scenarios for these planets can be explored to analyse the attenuation of the UV radiation. A recent study from \cite{OMalley2017} used different atmospheres to investigate the potential effect of UV radiation on surface habitability of the Trappist-1 HZ planets and found that oxygenic atmospheres are crucial in protecting the surface from the UVC, even for a thin-oxygen atmosphere (0.1 bar). However, this study did not consider the direct effects of powerful flares.

Here, we calculate the temporal evolution of the UV (100-450 nm) spectra for the most complex and energetic superflare observed in Trappist-1 with the K2 mission (Section ~\ref{sec:spec0}). This superflare released an energy of 1.24 $\times$ 10$^{33}$ erg within a total duration of 72 minutes. The structure of the superflare is very complex, with two impulsive phases: one at $\sim$ 1100s and a second one at $\sim$ 1850s, after the start of the flare. We use these spectra as an input in a radiative transfer code to analyse the variation of the surface UV radiation for each interval of the superflare. Two atmospheric scenarios are adopted for each planet to analyse the attenuation of the UV reaching the surface: a primitive (Archean) and a present-day atmosphere  (Section~\ref{sec:atmo}). The biological impact of the UV radiation is analysed for two bacteria ({\it D. radiodurans} and \textit{E. coli}) present in the surface or in the ocean of the planets. The results are shown in Section \ref{sec:results} and the main conclusions are presented in the last section.




\section{Methods}
\subsection{Estimation of approximate UV Spectra of Trappist-1 superflare}
\label{sec:spec0}

The observation of the strongest known flare or so called `superflare' of Trappist-1, that we consider for our study is described in \cite{Vida2017}. They provide integrated lightcurve of the complex flare in the wavelength range of 430-890 $nm$, obtained from K2 observation. This integrated lightcurve is in terms of differential magnitude (${\Delta} \rm mag$), where the first as well as the biggest peak of the complex flare profile has a magnitude of 1.78. To obtain the mean lightcurve in a given wavelength from the values of ${\Delta} \rm mag$, we use the following equation described by \cite{Vida2017}:

\begin{equation}
\label{eq:inten}
    \frac{I_{0+f}(t)}{I_{0}} = 10^{\frac{\Delta \rm mag}{2.5}}
\end{equation}
\noindent where, $I_{0+f}$(t) is the flare intensity as a function of time and $I_{0}$ is the mean quiescent flux or intensity of the star as seen from the Earth. We estimate $I_{0}$ of the Trappist-1 star using the Phoenix BT-Settle model (\cite{Allard2012}), which is a theoretical cloud model for stellar spectra, that can reproduce spectra of main sequence stars down to the L-type brown dwarf regime. It allows the input of three main parameters, namely, the effective temperature ($T_{\rm eff}$) in Kelvin, log gravity (in $cm s^{-2}$) and metallicity of the star. For Trappist-1, we use as input for those parameters the following values: 2511 K, 5.227 and 0.04, respectively.
We normalize the resulting spectrum to obtain the quite time spectrum that should be seen from Earth and then compute $I_0$ in K2 wavelength range ($\Delta \lambda_{K2}$) with Equation~\ref{eq:idel}.

\begin{equation}
\label{eq:idel}
  I_0 \Delta \lambda_{K2} = \int^{K2} 4 \pi {R_T}^2 F_{\lambda} d\lambda
\end{equation}

The calculated mean flux, $I_{0+f}$(t) of the complex Trappist-1 superflare in the K2 wavelength range is shown in Fig. \ref{fig:lightcurve} (left) . We use this lightcurve to find the approximate spectra of the flare, as described in the next two following steps.

\subsubsection{Step 1: Finding the template spectrum and flux values from available AD Leo flare data}
\label{sec:spec1}

The detailed spectroscopic observations of stellar flares in M-Dwarf stars are rare. Stellar flares are produced due to same (not well understood) physical process in all cases, and so, they should have general trends for their spectral evolution. With the knowledge of lightcurve evolution of the Trappist-1 flare in various spectral ranges, we used the spectral and lightcurve information of another observed superflare in the M dwarf star AD Leo presented by \cite{Hawley1991} to find the approximate spectra of the Trappist-1 flare, obtained by systematic extrapolation following the methods of \cite{Segura2010}. Here, the comparison of lightcurves of the Trappist-1 flare with those for the AD Leo one shows that they are not of the same order (even considering the relative distance of those from Earth). Yet we proceed with the assumption that the correlation of the spectral evolution of a strong flare (superflare) with that of mean lightcurve in various wavelength ranges in the UV-visible range for the M-dwarf star AD Leo also apply for the superflare of the Trappist-1, considered here, so that the Trappist-1 spectral evolution may be considered as a scaled down version of that in the AD Leo flare. The method deviates in one aspect from that of \cite{Segura2010}, in the fact that suitable scaling is also introduced during the calculation. The assumption stated above may seems a crude one, but considering the fact that we are not aiming for the minute analysis of spectral and temporal behavior of the Trappist-1 flare and its underlying physics, rather interested in estimating the gross influence of the flare on the surface environment and habitability only, the approximation is valid and crucial. \cite{Hawley1991} followed Johnson Photometric system (\cite{Jon95}) to calibrate the visible part of the spectra in U,B,V and R and presented the mean lightcurve (per unit wavelength in the range) of the AD Leo flare in their paper. For our convenience of calculation, following the usual method of subdivision (into Near, Middle and Far-UV) of the ultraviolet range of spectrum and the above convention for the visible part we divide the whole range of UV-Optical wavelength in 6 parts, as described in Table 1. Note that the U part of the spectrum (with a mean wavelength at about 365 nm) corresponding to the Jhonson Photometric system almost resembles that of the Near Ultra Violet (NUV) part of the conventional UV classification.

\begin{table}[!ht]
  \begin{center}
  \caption{}
  \label{tab1}
 {\scriptsize
  \begin{tabular}{|c|c|}\hline
 {\bf Division name} & {\bf Wavelength range ($nm$)} \\
 \hline
 UVA & 315 - 400 \\
 \hline
 UVB & 280 - 315 \\
 \hline
 UVC & 100 - 280 \\
 \hline
 FUV & 120 - 200 \\
  \hline
 MUV & 200 - 300 \\
  \hline
 U or N(UV) & $\sim$ 300 - 400   \\
\hline
 B & $\sim$ 400 - 500   \\
 \hline
 V & $\sim$ 500 - 600   \\
\hline
 R & $\sim$ 600 - 720  \\
\hline
\end{tabular}
  } \label{tab:res}
 \end{center}
\end{table}

Our very first interest has been to approximate spectra of the AD Leo flare of \cite{Hawley1991} at few different times over the course of the whole flare, where very limited number of disjoint observed spectra at three different ranges are presented in Figure 4 (N(UV) + part Optical), 5 (MUV) and 6 (FUV) of their paper. To calculate the complete spectra over the whole time period of the AD Leo flare, which consists of two peaks, we follow the method described by \cite{Segura2010}. For the simplification of the extrapolation method used, they approximated the double peaked flare as consisting of a single peak followed by a gradual decline, such that the spectrum at 915s represented the only peak of the flare. There are no spectral observations in the MUV ($\sim$ 200 - 300 $nm$) any time near that peak of the flare and the first such spectrum is observed only at 1604s after the flare starts. To find the approximate MUV part of the spectrum at the adjusted peak ($\sim$ 915 sec), the long wavelength end of the first MUV spectrum at 1604s, as presented in \cite{Hawley1991} (topmost picture of Figure 5a of their paper) is scaled by multiplication with a factor of $\sim$ 3.6 to combine it with the short wavelength end of the peak NUV+Optical spectrum at 915s (third picture from top of Figure 4a). Then we join it with the FUV spectrum at 900s (Figure 6 of \cite{Hawley1991}) to get the complete (FUV+MUV+N(UV) + part optical(B)) spectrum of the adjusted peak for the AD Leo flare. 

Now, as we have only the mean lightcurve of N(UV), B, V and R part of the whole AD Leo flare from \cite{Hawley1991}, we compute the other parts of the lightcurves, namely the MUV and FUV parts, of the same flare. To do this, we first normalize the N(UV) and B parts of the calculated peak flare spectrum using the above mentioned lightcurve of the same ranges to obtain the approximate AD Leo spectra at few different times during the flare. Then we scale the remaining (FUV + MUV) parts of the peak spectrum to match the N(UV),B part of the spectra at those times. Following, the integration of respective parts of the spectra over those two wavelength intervals (FUV and MUV) and dividing by the corresponding interval we get the mean light curves of the complex AD Leo flare in FUV and MUV also.

At this point we have the two parts of the AD Leo flare template, namely, the approximate peak spectrum at $\sim$ 915s and the approximate light curves of the same flare at different wavelength ranges (FUV + MUV + N(UV) + B + V + R).  

\subsubsection{Step 2: Finding the spectra of Trappist-1 flare using the templates}
\label{sec:spec2}
\begin{figure}[!ht]

\begin{center}
\includegraphics[width=6.5cm]{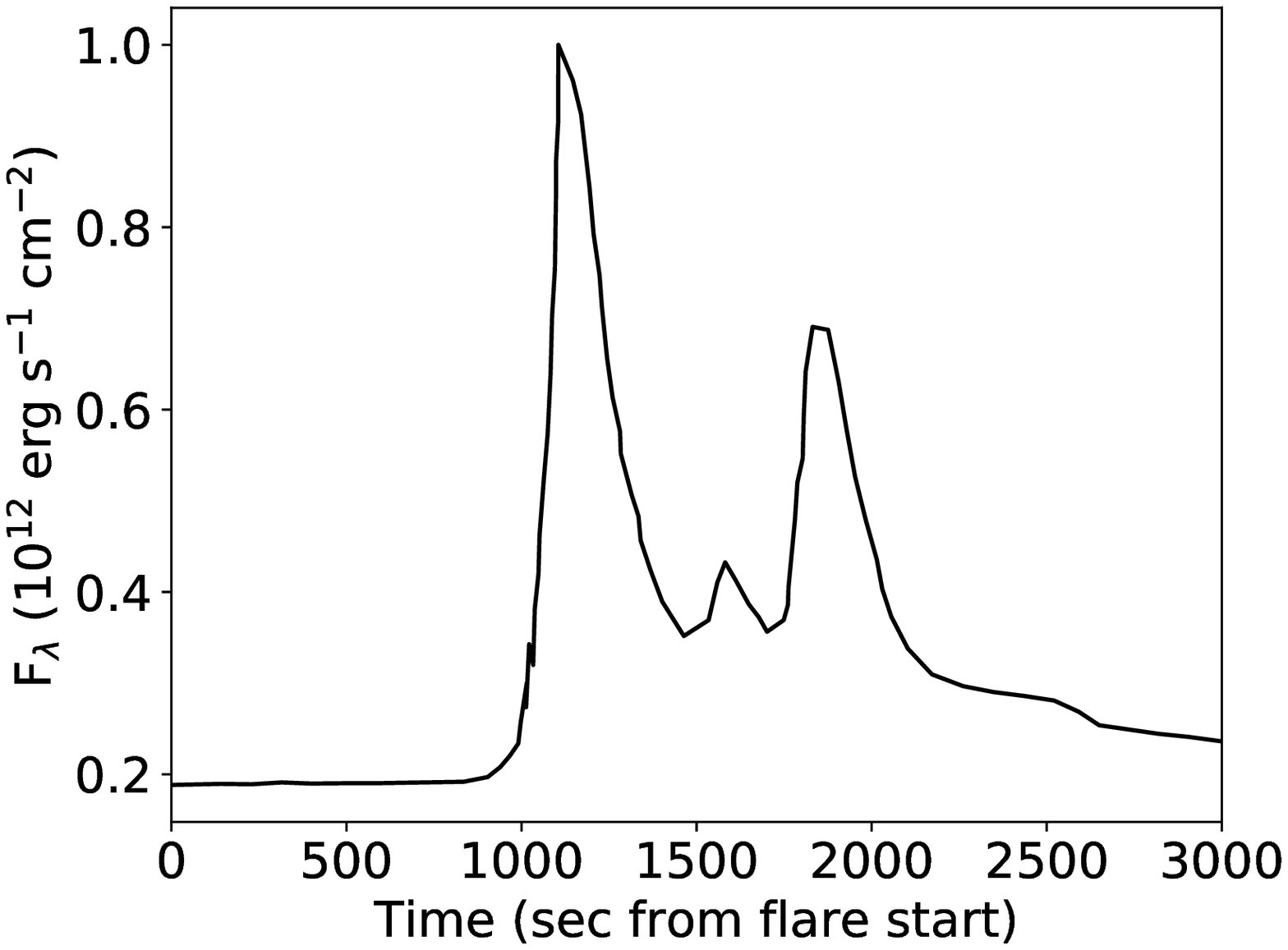}
 \includegraphics[width=8cm]{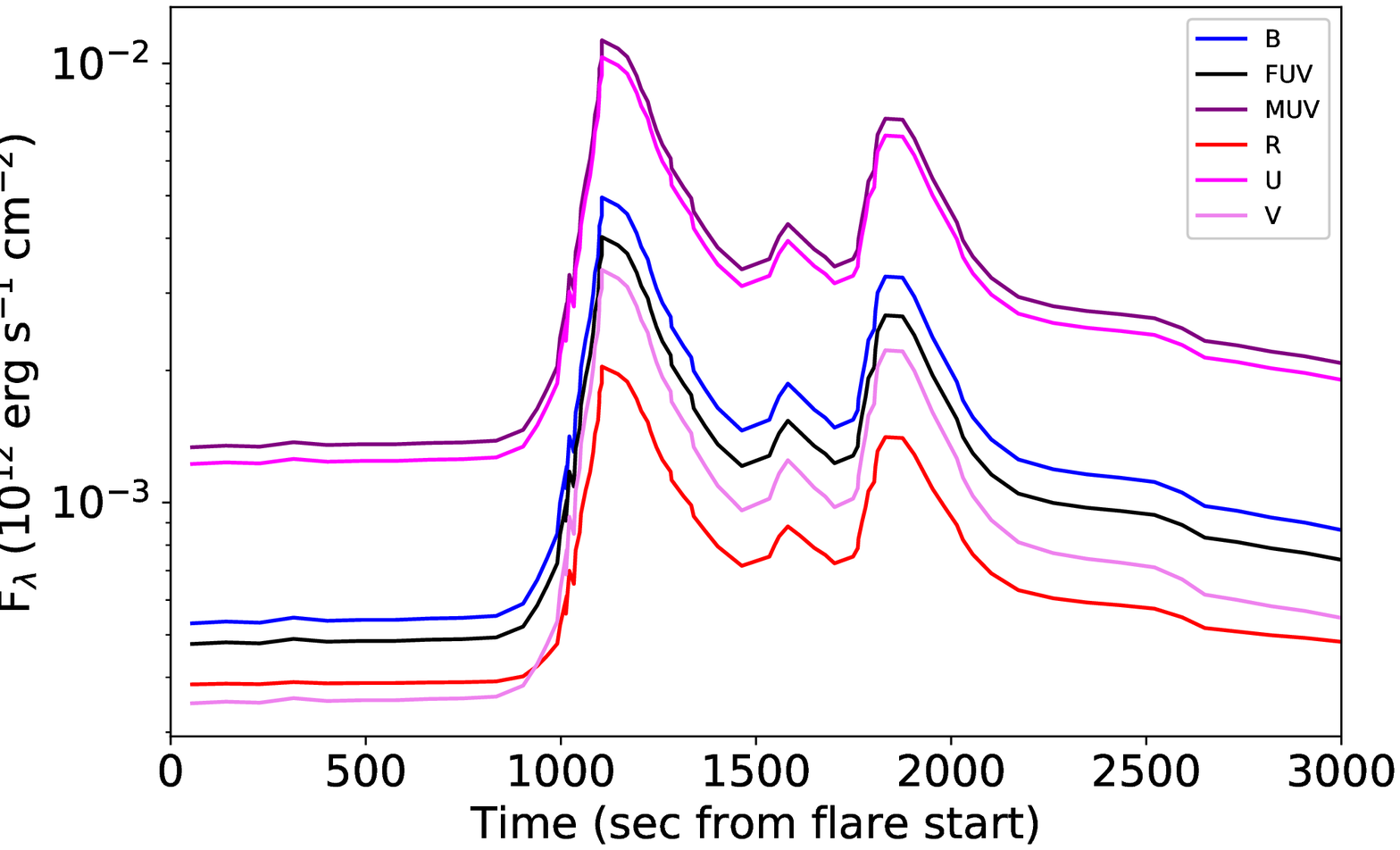}

 \caption{Approximate mean flux (left) of the Trappist-1 flare in K2 wavelength range (observed from Earth), calculated using equation~\ref{eq:eff} and estimated mean fluxes of the Trappist-1 flare in various wavelength ranges (right) are shown in the figure.}
   \label{fig:lightcurve}
\end{center}
\end{figure}

\begin{figure}[!ht]

\begin{center}
 \includegraphics[width=15cm]{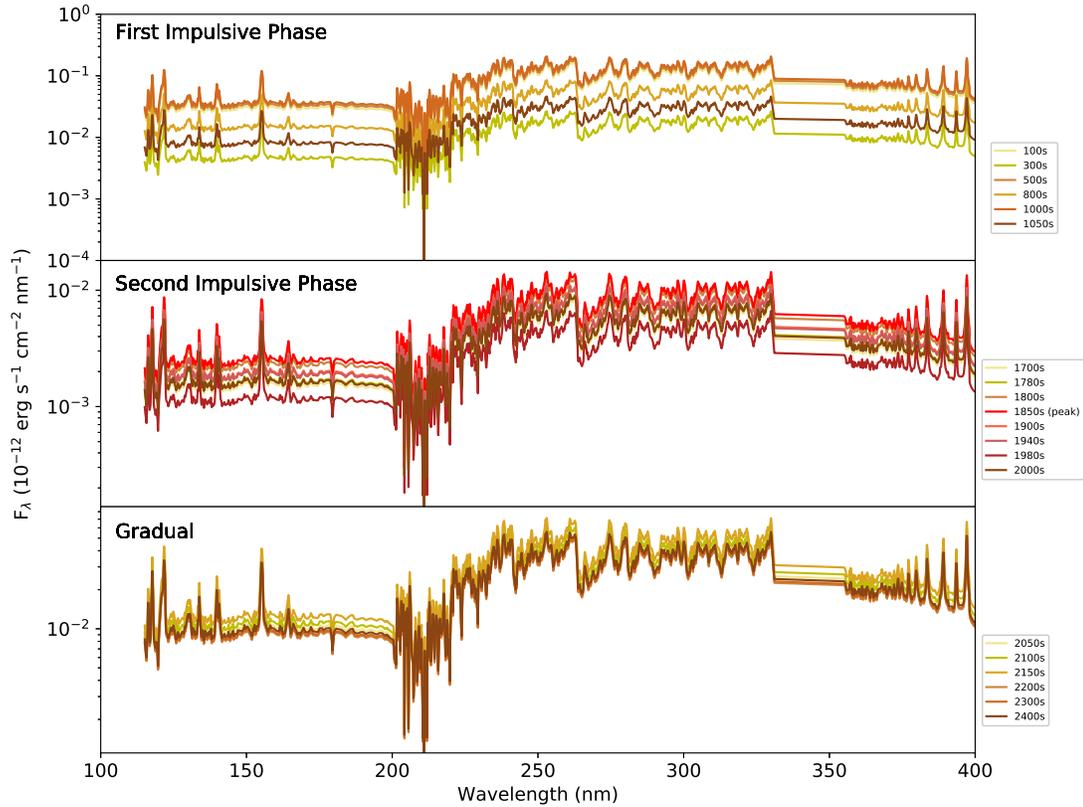}
 \caption{Estimated UV spectra of the strongest flare of Trappist-1 from \cite{Vida2017} for every 5 minutes during the evolution of the flare.}
   \label{fig:spectra}
\end{center}
\end{figure}

The assumption of applicability of the time correlation of the mean fluxes at different wavelength ranges of the flare of one M-dwarf star, namely AD Leo to a flare of another M-dwarf, i.e., Trappist-1 allows us to obtain the fluxes in all other wavelength ranges of the later. The spectral interval, out of those listed in Table~\ref{tab:res}, that should best approximate the K2 wavelength range in terms of mean lightcurve is the R(600-720 nm). Mean flux at all other wavelength ranges for the Trappist-1 flare are obtained by interpolation of those for AD Leo flare using the mean flux information in R ($\approx$ K2 range) of the former. The estimated values of the approximate mean fluxes of the Trappist-1 flare are shown in Figure \ref{fig:lightcurve} (right). Finally, scaling down and normalizing different parts of the template (approximate peak) spectrum of the AD Leo flare with the time varying mean flux values of Figure \ref{fig:lightcurve} and combining those resultant parts we find the approximate spectra (Figure \ref{fig:spectra}) of the Trappist-1 flare, that should be observed from Earth, throughout the time span of the flare.

\subsection{Atmospheric simulation} 
\label{sec:atmo}

To compute the radiative transfer of UV radiation (180-300 nm) through the atmospheres of the Trappist-1 planets, we used a two stream radiative transfer code from \cite{Ranjan2017}. This code computes the UV fluxes and intensities in the top of the atmosphere and at the surface of the planets orbiting M-dwarfs, under a specified atmosphere and surface conditions. It follows the treatment of \cite{Toon1989}, requiring the partinioning of the atmosphere into N homogeneous layers and the zenith angle of the star. In addition, the code also requires as input the temperature, pressure and composition (gas molar concentration) as a function of altitude. We partitioned the atmosphere into 55 layers, each having thickness of 1km, and two atmospheric scenarios were used as an input to the code: (i) a 1 bar CO$_{2}$ dominated atmosphere (0.9 bar N$_{2}$, 0.1 bar CO$_{2}$), similar to the Archean Earth at 3.9 Gyr and (ii) a modern atmosphere with ozone. For the former, we adopted a pre-biotic model already provided by the code. While for the latter, we consider an atmosphere composed of N${_2}$, O$_{2}$, CO$_{2}$, H$_{2}$O, CH$_{4}$, O$_{3}$, and SO$_{2}$, and we build our own model according to the following steps:
\begin{enumerate}
\item[1)] We obtained the mixing ratio of the trace gases like H$_{2}$O, CH$_{4}$ and N$_{2}$ for Trappist-1 e, f and g from \cite{OMalley2017}. We also used their constant and non-negligible mixing ratio for N$_{2}$O of 1.5 $\times$ 10$^{-2}$.
\item[2)] In the study from \cite{Hu2012}, the mixing ratios of CO$_{2}$ are found to be constant at all heights, due to the fact that CO$_{2}$ is well mixed in the atmosphere. Considering this fact, we assumed a constant mixing ratio of CO$_{2}$ of $\sim$ 370ppm (0.00037 mole/mole), like the one of the current Earth (\cite{Foucher2011}). However, previous studies \citep{Linc18,Turbet18, Hu20} suggest that Trappist-1 f and g could require abundance of CO$_{2}$ to be habitable.
\item[3)] We adopted the concentrations of O$_{2}$ and N$_{2}$ from the MSIS-E-90\footnote{\url{https://ccmc.gsfc.nasa.gov/modelweb/models/msis_vitmo.php}} model for a current Earth. This model gives the concentrations of O, N$_{2}$, O$_{2}$, He, Ar, H and N. However, we excluded the contribution of He, Ar, H as the contribution of these species in the total mass density is very small (around 10$^{-2}$- 10$^{-3}$) with respect to O$_{2}$ and N$_{2}$ densities. Therefore, we calculate the mixing ratios of N$_{2}$, O$_{2}$, O and N by taking the ratio of concentration of individual species with respect to the total (summed) concentration. Finally, we add the mixing ratios of N and O to those of O$_{2}$ and N$_{2}$.
\item[4)] For the mixing ratio of SO$_{2}$, we use the concentration from \cite{Rug15} for a Earth-like planet orbiting a M8V star (same spectral type as Trappist-1) with an age of 3.9 Gyr. These data are available in the code.

\item[5)] Finally, to build the Temperature-Pressure-Density profile, we used the temperature as a function of height from \cite{OMalley2017} for a 1 bar atmosphere and we adopted the same pressure as function of height as that of the current Earth. To obtain the total concentration (molecules cm$^{-3}$) of the air as function of height, we have to calculate the concentration of all those species from their mixing ratio values and add them. It is important to note that as the process involves data from different sources, the final results are very approximate.
\end{enumerate}

\begin{figure}[!ht]

\begin{center}
\includegraphics[width=10cm]{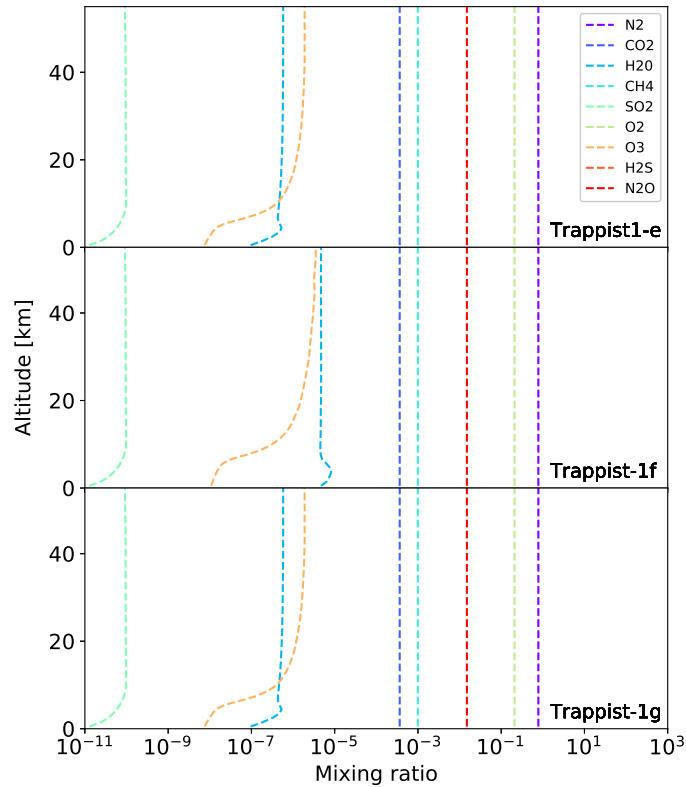}
 \caption{Mixing ratios of the trace gases used as input in the model for the three planets in the HZ of TRAPPIST-1.}
   \label{fig:mix}
\end{center}
\end{figure}

The final mixing ratio of the gases for a modern Earth atmosphere is shown in Fig. \ref{fig:mix} for each of the Trappist-1 HZ planets.

\subsection{Biological Impact}
\label{sec:life}
Superflares significantly increase the UV flux of the star, and consequently can affect the conditions for the origin and development of life. Shorter UV wavelengths (10-200 nm) can be absorbed in the top of the atmosphere if the planet has strong absorbers like N$_{2}$, CO$_{2}$ or H$_{2}$O. While UV wavelengths within 200-300 nm (MUV) and 300-400 nm (NUV) can partially reach the surface of the planet depending whether the planet possess an ozone layer or not. The DNA molecules of living beings are mainly damaged by the UVC and UVB range. Therefore, here we focus on the probable impact that the increase of the MUV+NUV+FUV radiation during the impulsive phase of a superflare in Trappist-1 could have on life present in either the surface or in the ocean of the potentially habitable planets Trappist-1 $e$, $f$, and $g$.

\begin{figure}[h]
\begin{center}
 \includegraphics[width=6cm]{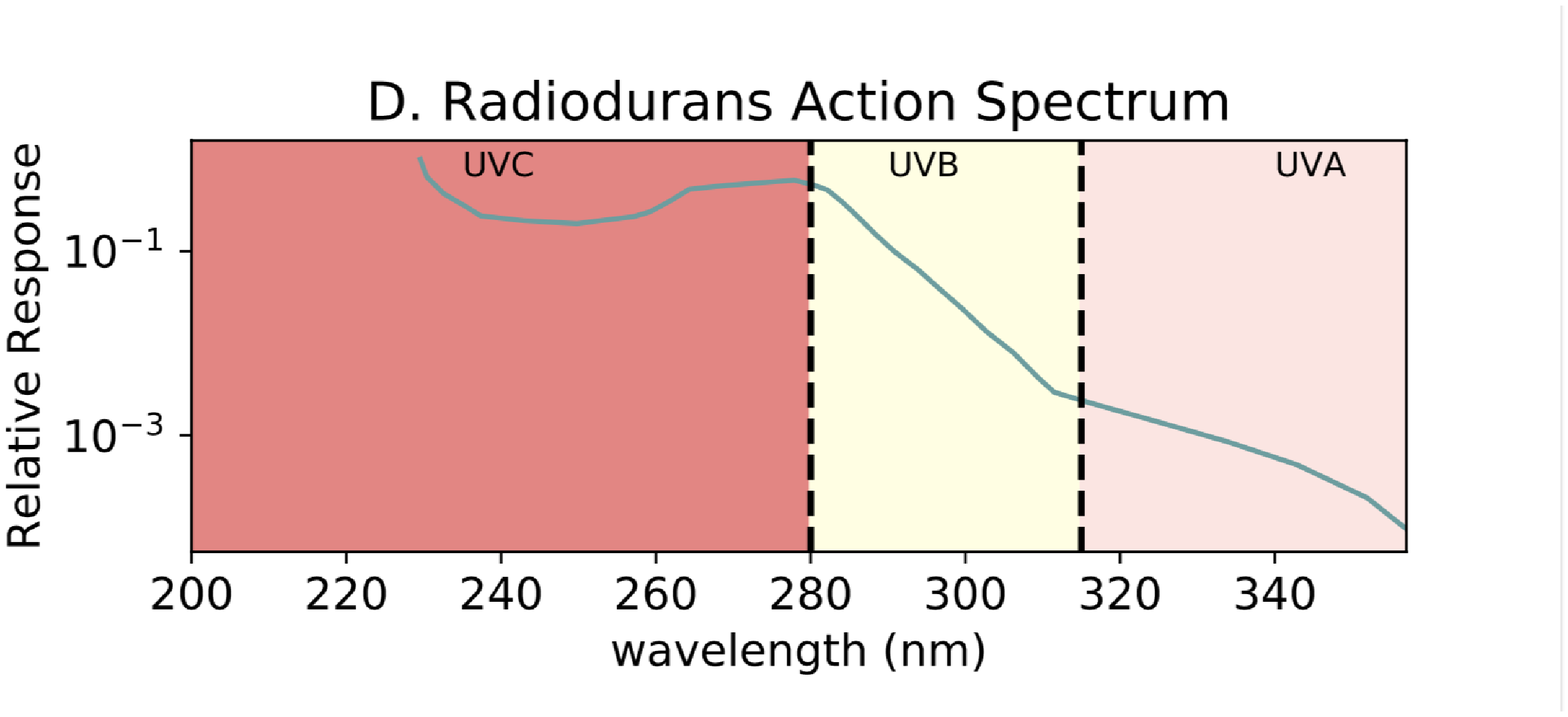} 
  \includegraphics[width=6cm]{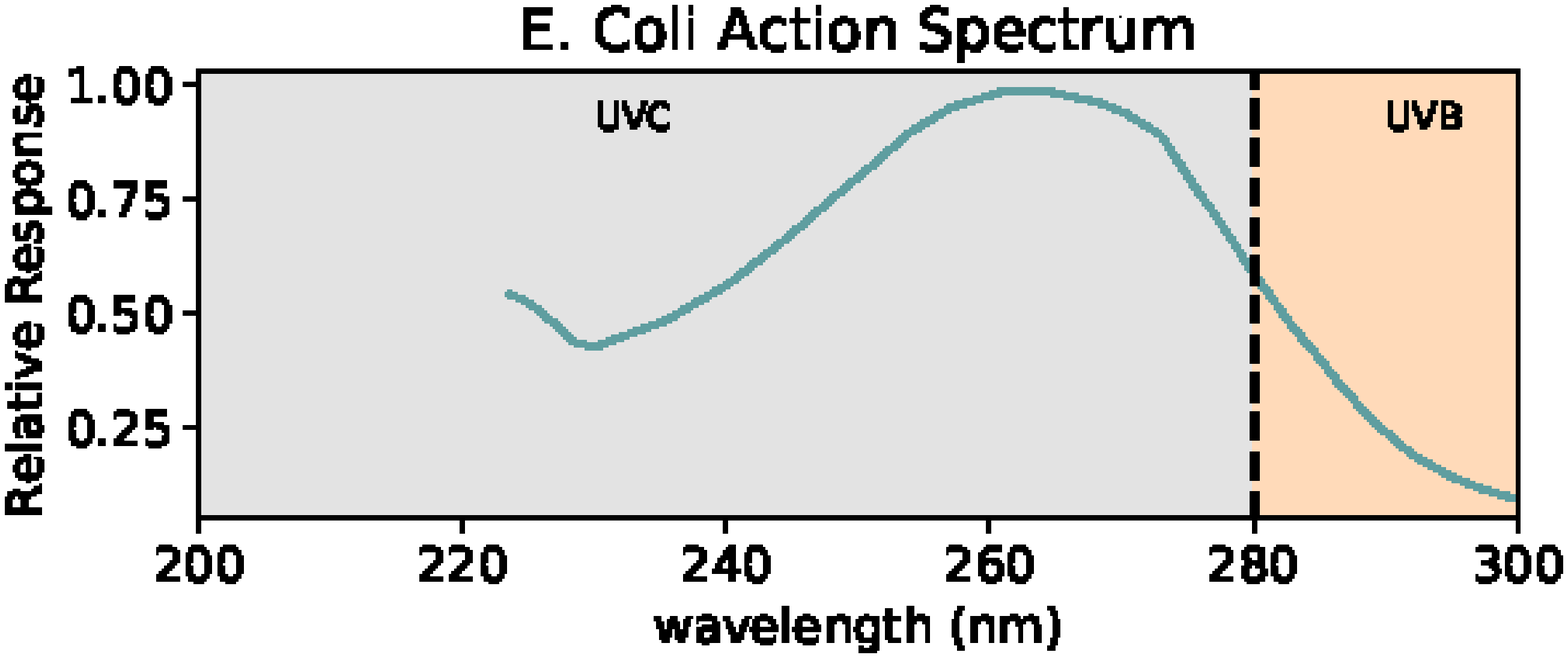}
 \caption{Action spectra, or biological response, for {\it E. coli} (left) and {\it D. radiodurans} (right) as shown in \cite{Estrela2018}.}
   \label{fig:actionspec}
\end{center}
\end{figure}

For this study, we chose two bacteria, a very resistant one, {\it Deinococcus radiodurans}, and a more common one, {\it Escherichia coli}. Since the response of a biological body varies as function of the wavelength, the effective biological response at different wavelengths has to be estimated. For that, the incident UV flux, $F_{UV}(\lambda$), is weighted by the action spectrum, $S(\lambda)$ of the microorganisms. 
\begin{equation}
    E_{\rm eff} = \int_{\lambda_{1}}^{\lambda_{2}} F_{\rm inc}(\lambda)\ S(\lambda)\ d\lambda
    \end{equation}\label{eq:eff}
\noindent where F$_{inc}$ is the total incident UV flux including the superflare  contribution arriving at the planet surface/ocean, $S$ is the action spectra, and $\lambda$ represent the middle-UV wavelengths (MUV, 200-300nm). The action spectra of the two bacteria are shown in Figure~\ref{fig:actionspec}. The UV dosage for 10\% survival of {\it D. radiodurans} is 553 J/m${^2}$, which corresponds to a UV flux (255 nm) of 1.7 W/m${^2}$ during 5 minutes \citep{Ghosal2005}, while for {\it E. coli} it is 22 J/m${^2}$.

To compute the incident UV flux (F$_{inc}$) at various depths $z$ of an ocean, we use the following equation:
\begin{equation}
I(\lambda, z) = I_{0}(\lambda) e^{-K(\lambda),z}
\end{equation}
\noindent where $I(\lambda, z)$ is the UV spectral irradiance at depth $z$, $I_{0}(\lambda)$ is the UV spectral irradiance with the superflare contribution passing through an Archean atmosphere and reaching the water surface, and $K(\lambda)$ is the diffuse attenuation coefficient for water given by the sum of the absorption coefficient of water and the scattering coefficient. For the details we would like to refer to \cite{Estrela2018}, in which the same methodology is applied to compute the biological impact by superflares from the solar-type star Kepler-96.

\vspace*{-0.1 cm}
\section{Results}
\label{sec:results}

\subsection{UV surface radiation during the evolution of the superflare}

\begin{figure}[t]
\begin{center}
 \includegraphics[width=15cm]{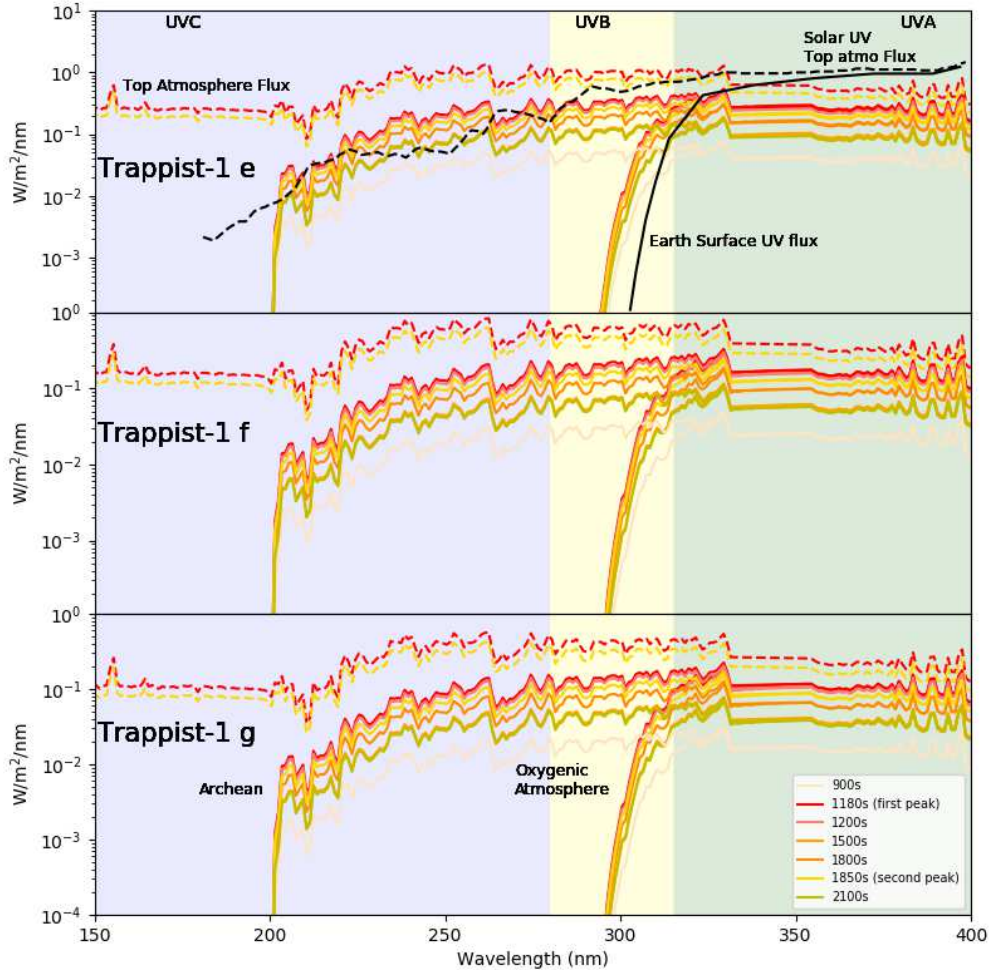}
 \caption{Ultraviolet flux at the top of the atmosphere (dashed line) and at the surface (solid line) of planets Trappist-1 e (top), Trappist-1 g (middle) and Trappist-1 f (bottom) during the evolution of the flare. The UV flux at the surface is transmitted by two atmosphere models: 1 bar CO$_{2}$ dominated atmosphere (Archean) and a present day Earth-like atmosphere with ozone. The quiescent solar UV flux received at the top of the atmosphere of Earth and the UV flux received at Earth's surface are plotted for comparison (dashed and solid black line, respectively).}
   \label{fig:surfflux}
\end{center}
\end{figure}

The net UV flux at the top of the atmosphere (dashed lines) and on the surface (solid lines) of the three Trappist-1's HZ planets are shown in Fig.~\ref{fig:surfflux} at intervals of 5 minutes during the occurrence of the superflare. As expected, the flux reaches higher values during the first impulsive phase ($\sim$ 1 Wm$^{-2}$nm$^{-1}$), but has a considerable increase during the second peak of the flare. On the other hand, our results show that due to the increase in UV by the superflare, the UV flux at the top of the atmosphere is 100 times higher than those found by \cite{OMalley2017} and  \cite{OMalley2019} for an active Trappist-1. These works use the spectrum of a flaring star based on the MUSCLES survey and Phoenix models, however they do not specify the energy range of the flare. It is also worth noting that a modern atmosphere with ozone absorbs all the radiation shortwards of 280 nm, which are the most dangerous to life, allowing just a small amount of UVB radiation to pass. The UVC and UVB in the TOA of Trappist-1 planets is slightly higher than the TOA of the present-day Earth (dashed black line in Fig. \ref{fig:surfflux}), while the UVA fluxes are lower. For an Archean atmosphere, only UV wavelengths smaller than 200 nm are absorbed, which means that the planetary surfaces still partially receive some UVB and UVA.

\subsection{Impact on life on the surface}
\label{surface}

Following \cite{Estrela2018}, we determine the survival of two bacteria on the surface of the Trappist-1's HZ planets by calculating the overall effective UV flux (E$_{\rm eff}$) that falls in a biological body (see Section~\ref{sec:life}), considering the strongest observed flare from \cite{Vida2017}. The results for the E$_{\rm eff}$ during the two impulsive phases of the flare are listed in Table~\ref{tab:res}. The threshold for the E$_{\rm eff}$ was chosen using the maximum UV flux for 10$\%$ survival of these bacteria.

We find that under an Archean atmosphere both bacteria would not be able to survive in planet-e during the first impulsive phase, but due to the UV resistance of \textit{D. radiodurans}, this bacteria could survive in any of the other planets. In contrast, the scenario is better for a planet with a present-day atmosphere (with ozone), for which, only during the first impulsive phase, \textit{E. coli} could not survive in Trappist-1 e. However, both bacteria could survive in the other HZ planets under the presence of an ozone layer. It is worth highlighting here that we are analysing the survival of Earth-like bacteria in a planet with a clear sky, however clouds or hazes could possibly absorb this UV radiation, and/or other bacteria could have evolved to survive under harsher UV conditions.

\begin{table}
  \begin{center}
  \caption{Biological effective irradiance, $E_{\rm eff}$ ($J/m^2$),
due to the two impulsive phases of the superflare. To obtain the values in Joules, we multiplied the values in Watts by the total duration of the each peak (146s and 158s, respectively).}
  \label{tab2}
 {\scriptsize
  \begin{tabular}{|c|l|c|c|c|}\hline
  \multicolumn{4}{|c|}{\bf First Impulsive Phase} \\
 \hline
{\bf Planet} & {\bf Bacteria} & {\bf Archean} & {\bf Present day} \\
 \hline
Trappist-1e &  {\it E. coli} & 1450  &  0.064  \\
 & {\it D. radiodurans} & 721 & 0.020 \\
  \hline
Trappist-1f & {\it E. coli} & 888 & 0.0152  \\
 & {\it D. radiodurans} & 440 &  0.0048     \\
  \hline
Trappist-1g & {\it E. coli} & 601 & 0.0132    \\
 & {\it D. radiodurans} & 298  & 0.0041   \\
\hline
  \multicolumn{4}{|c|}{\bf Second Impulsive Phase} \\
 \hline
{\bf Planet} & {\bf Bacteria} & {\bf Archean} & {\bf Present day} \\
 \hline
Trappist-1e &  {\it E. coli} & 1104  &  0.049     \\
 & {\it D. radiodurans} & 548 & 0.015     \\
  \hline
Trappist-1f & {\it E. coli} & 647 &  0.0016  \\
 & {\it D. radiodurans} & 335 &  0.0036     \\
  \hline
Trappist-1g & {\it E. coli} & 456 & 0.010     \\
 & {\it D. radiodurans} & 226  & 0.003   \\
\hline
  \end{tabular}
  } \label{tab:res}
 \end{center}
\end{table}

\subsection{Impact on life in the ocean}
\label{sec:ocean}
\cite{Estrela2018} showed that an ocean in a hypothetical Earth-like planet orbiting Kepler-96 could protect organisms from the increased UV radiation, allowing life in depths within the photic zone (up to 200m). Previous study from \cite{Grimm2018} found that the densities of the Trappist-1 planets range from 0.6 to 1 Earth's density, which implies that planet-e has largely rocky interior, while planets f and g require thick atmospheres, oceans or ice, with a water mass fraction of less than 5$\%$. Another study from \cite{Bourrier2017} used the instrument STIS from the Hubble Space Telescope to measure the ultraviolet radiation of Trappist-1 at Lyman-$\alpha$ and calculated the water loss rates in the planets. They found that the planets in the HZ should have lost less than three Earth-oceans of water, indicating that they could still have some remaining water.

\begin{figure}[t]
\begin{center}
 \includegraphics[width=15cm]{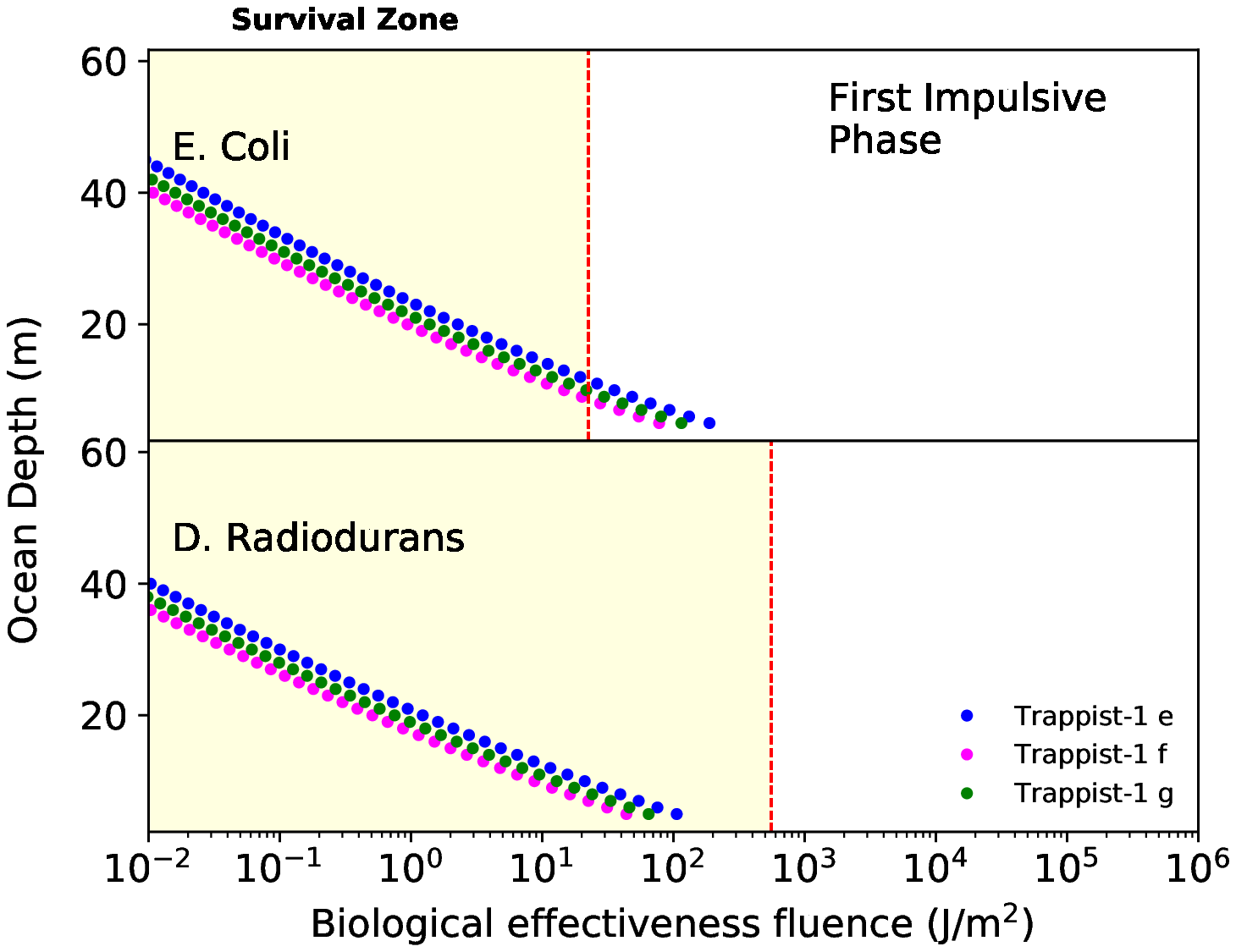}
 \includegraphics[width=15cm]{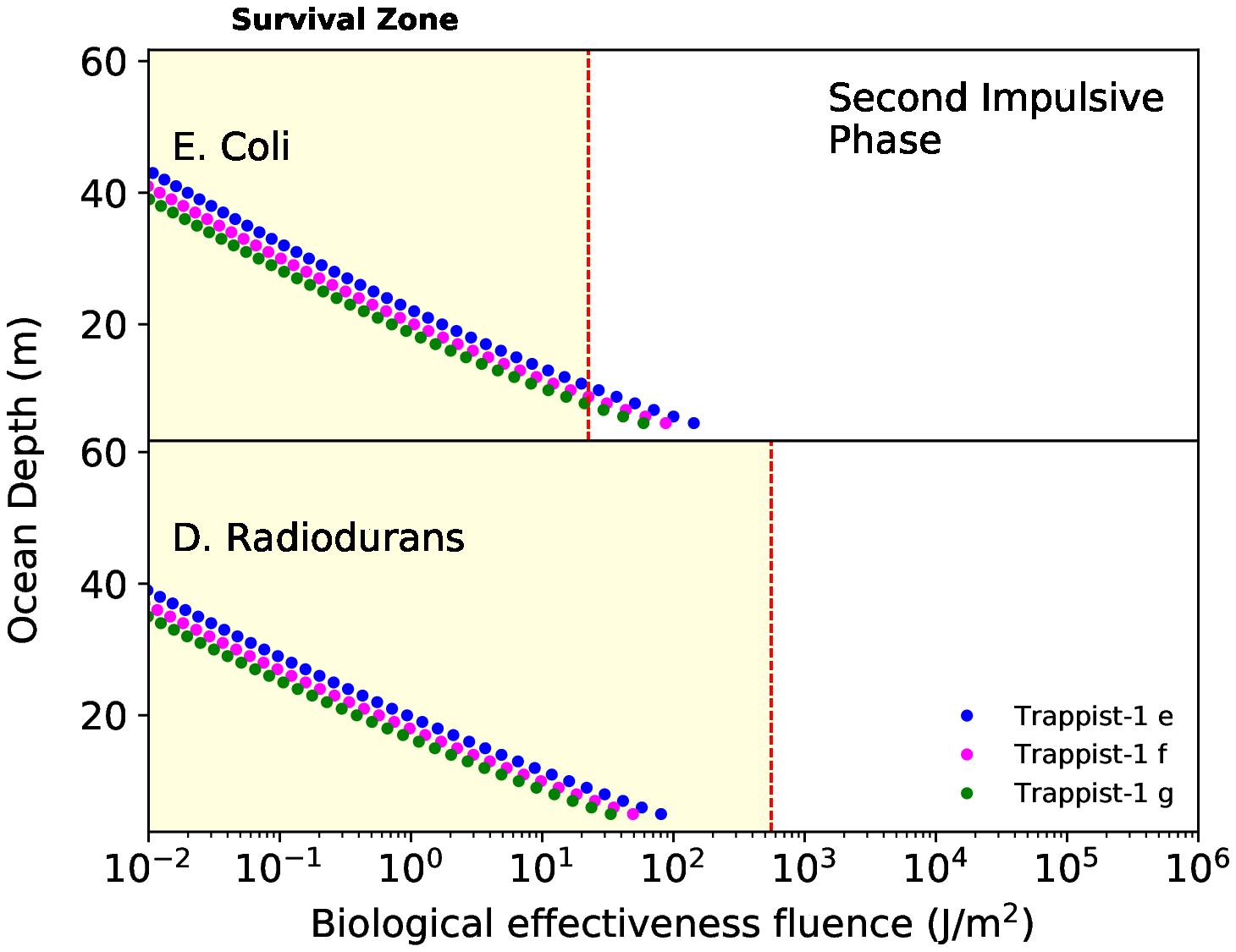}
 \caption{Biologically effective irradiance for \textit{E. coli} (E$_{\rm eff}$) with depth in an Archean ocean present in Trappist-1 e, f and g. The values in Joules of the E$_{\rm eff}$ were obtained by multiplying the values in Watts with the time duration of the two impulsive phases of the flare. The red vertical line represents the threshold for finding life determined by the maximum UV flux for 10\% survival of \textit{E. coli} and of \textit{D. radiodurans}.}
   \label{fig:ocean}
\end{center}
\end{figure}

Considering the possibility of the presence of an ocean on these planets, we analyse whether this could be a favourable condition in protecting the microorganisms (\textit{D. radiodurans} and \textit{E. coli}) from the increased UV flux due to superflares, under the lack of an ozone layer. 
We calculated the effective UV flux received by these bacteria at various ocean depths during the two impulsive phases of the flare, as described in Section \ref{sec:life}. The values of the UV irradiation varying with ocean depths are shown in Figure \ref{fig:ocean}. During the impulsive phases, \textit{D. radiodurans} could survive on the ocean surface. On the other hand,  \textit{E. coli} could survive approximately at 8, 9, and 11m below the ocean surface in the three HZ planets, respectively. These ocean depths are within the photic zone of $\sim$10m proposed by \cite{Heath99} for an M dwarf star with temperature similar to Trappist-1. \cite{Kiang07} also analysed the UV impact on life under the water based on the plant damage limit. They claim that flares of 10$^{33}$ ergs are at the threshold at which protection under water is necessary. However, they found that the safest ocean depth for a flare with this energy range is at 1.5m.

\section{Discussions}

\subsection{Impact of consecutive superflares}
In this work we show that the presence of an ozone layer is crucial in protecting life in the Trappist-1 HZ planets from the harmful UV flux of the strongest superflare observed from Trappist-1. However, we do not analyse disturbances in the chemical composition of the atmosphere of the planets by the Trappist-1 flare due to a single or consecutive flares. Previous works from \cite{Venot16} and \cite{Segura2010} focused on the effects that the great flare observed on the active M-dwarf AD Leo could have on the composition of the atmosphere of an Earth-like planet orbiting at 1AU of this star. \cite{Venot16} suggest that the atmosphere would return to its steady state after $\approx$ 30,000 years post-flare. Moreover \cite{Segura2010} show that the recovery of the atmosphere depends if the flare is associated with energetic particles. They show that the ozone layer recovers in about 30 hours if there are no energetic particles enhancement, otherwise the ozone layer could be depleted for about 30 years. 

More recently, \cite{Tilley19} analyzed the impact of consecutive flares on the ozone layer of an Earth-like planet, and found that single lower-energy flares do not significantly impact the ozone column. They also state that even under sequential (every 2 hours) superflares with 10$^{34}$ erg, the impact on the ozone column is lower than the impact by a single proton event associated with a flare in this energy range. However, if lower-energy flares ($3\times10^{30}$ erg) associated with proton events occur in a frequency higher than once per year,  the ozone column can rapidly be eroded.

In the case of Trappist-1, \cite{Vida2017} found a total of 42 flares, and only one of them reach 10$^{33}$ erg, while the others have energy varying between 10$^{30}-$10$^{32}$ erg. These flares occur in a mean interval of 28 hours and it is not known if they are associated with proton events. If they are, according to the results from \cite{Tilley19}, a 1 day separation between flares of $8\times10^{31}$ erg would deplete the ozone layer for 2.3 years. The removal of the ozone layer would allow the irradiation of harmful UV wavelengths, making the surface of these planets a hostile place for life. Therefore, by not taking into account the UV increase due to consecutive flares and its impact on the ozone layer, our study is underestimating the UV radiation and its effects on life.

A strong magnetic field in these exoplanets could also shield the atmosphere from the impacts of the enhanced proton flux associated with such superflares. However, according to \cite{Vidotto13}, to possess a magnetosphere that is able to protect the atmosphere from these eruptions, Trappist-1 HZ planets would need to have strong magnetic fields of order $\approx$ 10-10$^{3}$ Gauss due to the high magnetic pressure of its host star, whereas Earth's magnetosphere is only about 0.5 Gauss.

All these factors can impose difficulties to the development of life in the surface of Trappist-1 planets. Therefore, we also analyzed here the possibility of life in an ocean on these planets, in the absence of an ozone layer. As described in Section~\ref{sec:ocean}, previous studies indicate the possibility of an ocean in the HZ planets of Trappist-1, which in turn, could protect life from the harmful effects of the increased UV by the superflares.

\subsection{Bacteria recovery between two impulsive phases}
\cite{Patel2009} irradiated cultures of \textit{D. radiodurans} with UV radiation (200-500 nm) for 5 min using a flux of 300 W/m$^{2}$ (1.2 W/m$^{2}$/nm). Then, they measured the growth rate curve of the bacteria during 24 h by measuring the number of colonies after the exposure to the UV irradiation. Their growth curve shows that a population of wild-type \textit{D. radiodurans} starts growing exponentially after $\sim$ 8 hours and in about 23 hours the population is completely recovered. The interval between the two main impulsive phases of the superflare is 14 minutes, and also the fluxes per nm are 100 times higher than the flux used by \cite{Patel2009}. This suggests that this bacteria in the surface of a planet with an ozone layer would not have sufficient time to recover between the first and the second impulsive phases of the flare. However, the bacteria could probably have time to recover before the arrival of a second flare in Trappist-1, as the interval between consecutive flares is 28 hours. We would like to emphasize that this recovery would  apply only to an Earth-like evolution that produced \textit{D. radiodurans}. Other organisms on that planet may have evolved faster regeneration capacities. \\


\subsection{Superflares and the origin of life}
In Sections \ref{surface} and \ref{sec:ocean}, we analysed how superflares could be lethal for already evolved life forms. However, previous studies (\cite{Ranjan2017}, \cite{Air16}) suggested that high energy flares could help the prebiotic chemistry. UV radiation is a key ingredient for several prebiotic reactions, however M dwarf stars emit lower in NUV than solar-type stars. If the prebiotic reactions that are UV dependent could occur at low irradiance levels of UV light, the reactions rates would be very small and the origin of life could be delayed (\cite{Ranjan2017}). M dwarf flares, such as the one analysed in this work for Trappist-1 and the AD Leo Great flare of 1985, can increase the NUV radiation at the surface of the orbiting planets up to 10$^{4}$ W/m$^{2}$/nm. Depending on the frequency of these flares, they could compensate the low NUV flux of the M dwarfs. Trappist-1 in particular has a high frequency flaring rate, with $\sim$0.75 cumulative flares per day. However, laboratory studies are necessary to determine if the frequency and duration of these flares could provide sufficient fluence to trigger the pre-biotic chemistry. 



\section{Summary and Conclusions}

The discovery of frequent energetic flares in the Trappist-1 by \cite{Vida2017} brings into question the habitability of the terrestrial planets of this system. In this work we calculate the UV (100-450 nm) spectra during the temporal evolution of the most energetic flare observed in Trappist-1 (1.24 $\times$ 10$^{33}$ ergs) using the well observed flare of the M4V star Ad Leo \cite{Hawley1991} as a template. Moreover, we use two atmospheric models, a primitive (Archean) atmosphere and an atmosphere with ozone, to analyze the attenuated incident UV flux during the interval of the flare and the impact that it could have in life present at the surface or in the ocean of Trappist-1 HZ planets. Our findings are:

\begin{itemize}
\item The superflare observed in Trappist-1 can increase the top of the atmosphere flux up to $\sim$ 1 Wm$^{-2}$nm$^{-1}$ in the first impulsive phase. The UVC fluxes can be up to 2 times higher than the one received at TOA of present-day Earth, but the UVA fluxes are lower.

\item The increase in the UV fluxes due to the flare makes the survival of the bacteria harder in planet-e under a primitive scenario, but UV resistant microorganisms, such as \textit{D. radiodurans}, are able to survive the irradiance received in planets f ang g in the first and second impulsive phases of the flare. On the other hand, the presence of an ozone layer absorbs most of the harmful UV radiation allowing bacterial lifeforms to survive at the surface under any atmospheric scenario.

\item An ocean in these planets could also provide a safe refuge for the lifeforms under the high UV irradiation of the flares. In the case of the bacteria analyzed in this work,  {\it D. radiodurans} could survive on the ocean surface, while {\it E. coli} at about 10m below the ocean surface of the three HZ planets during the first impulsive phase of the superflare.

\end{itemize}

\acknowledgments

\section{Acknowledgements}
We would like to thank Jean Pierre Raulin for the encouragement in the preparation of this manuscript. Raissa Estrela acknowledges a FAPESP fellowship ($\#$2016/25901-9 and $\#$2018/09984-7). A. Valio acknowledges partial support from FAPESP grant (\#2013/09824-6). This research was carried out at Center for Radioastronomy and Astrophysics Mackenzie and at Jet Propulsion Laboratory, California Institute of Technology, under a contract with the National Aeronautics and Space Administration.

\section{Disclosure statement}
No competing financial interests exist.

\end{document}